\documentclass[12pt]{iopart}

\usepackage[T1]{fontenc}
\usepackage[utf8]{inputenc}

\expandafter\let\csname equation*\endcsname\relax
\expandafter\let\csname endequation*\endcsname\relax

\usepackage{amsmath,amssymb,mathtools,graphicx,amsthm}
\usepackage{bm,soul}
\usepackage{hyperref}
\usepackage{dsfont,bbm}
\usepackage{xcolor}
\usepackage{enumitem}
\usepackage{float} 
\usepackage{caption}
\usepackage{cancel,ulem}

\usepackage{tikz}
\usepackage{pgfplots}
\usetikzlibrary{decorations.pathmorphing}
\usepackage[scr=boondoxo,scrscaled=1.05]{mathalfa}
\hypersetup{
	colorlinks  = true, 
	urlcolor   = darkgreen, 
	linkcolor  = darkblue, 
	citecolor  = darkred 
}
\usepackage{xcolor}
\colorlet{darkred}{red!85!black}
\colorlet{darkgreen}{green!50!black}
\colorlet{darkblue}{blue!60!black}
\definecolor{darkorange}{rgb}{1.0, 0.55, 0.0}
\definecolor{darkspringgreen}{rgb}{0.09, 0.45, 0.27}
\newcommand{\tf}{t_{\mathrm{f}}}
\begin{document}

\title{What is a minimum work transition in stochastic thermodynamics?}

\author{Paolo Muratore-Ginanneschi and Julia Sanders}

\address{University of Helsinki, Department of Mathematics and
	Statistics P.O. Box 68 FIN-00014, Helsinki, Finland}

\eads{\href{mailto:paolo.muratore-ginanneschi@helsinki.fi}{paolo.muratore-ginanneschi@helsinki.fi}, \href{mailto:julia.sanders@helsinki.fi}{julia.sanders@helsinki.fi}}

\begin{abstract}
  We reassess the concept of transition at minimum work in classical stochastic finite-time thermodynamics, when the system dynamics is modelled by a diffusion process. We show that a well-posed formulation of the optimal control problem corresponding to the minimisation of the mean work done on the system during a finite-time transition necessarily requires taking into account speed limits on control protocols. This fact has major qualitative consequences. First, it permits to discriminate between optimal swift engineered equilibration and transitions at minimum work. Second, it shows that in the limit when speed limits are removed, only transitions specified by generalized Schr\"odinger bridges admit a consistent physical interpretation. To illustrate these points, we focus on the simplest model problem: a particle in a viscous fluid confined in a Gaussian moving trap. 
\end{abstract}


{
	\let\newpage\relax  
	\maketitle
}

\section{Introduction}
The operational principles of {\textquotedblleft}small{\textquotedblright}, i.e. micro or nano, systems differ from those of macroscopic systems. 
Small systems constantly experience random thermal and topological fluctuations that may drown out any mechanical behaviour \cite{KaLeZe2007}. Mathematical models based on stochastic differential equations are therefore more adapted to describe small system dynamics over finite time-horizons \cite{PePi2020}. In particular, these models capture a central feature of small systems: in order to be functional, small systems need to operate out of thermal equilibrium. This is because equilibrium forces produce net zero probability flux, as a consequence of detailed balance \cite{BrSi2019}. 

In this context, stochastic optimal control theory \cite{BecJ2021,BoSiSu2021} can be used to investigate how small systems can efficiently harness randomness in order to generate controlled motion or perform thermodynamic work on larger scales. Indeed, insights from optimal control paved the way for experimental tests of the second law of thermodynamics at the small scale \cite{BeArPeCiDiLu2012}, and the laboratory realization of sub-micro scale Stirling engines \cite{BlBe2011,KoMaPeAv2014,MaPeGuTrCi2016,MaRoDiPePaRi2016}. The theoretical analysis of minimum mean work transitions conducted by Schmiedl and Seifert in \cite{ScSe2007} played a groundbreaking role towards these applications, alongside other theoretical work, see, e.g., the recent review \cite{GuJaPlPrTr2023}. 

Schmiedl and Seifert already warned in \cite{ScSe2008} that the mathematical formulation of the minimum work transition in \cite{ScSe2007} ``\emph{can lead to unphysical optimal processes which are driven infinitely fast with infinitesimally small changes [of the state variable]}''. As a consequence, to make predictions that can be compared with experiments, Schmiedl and Seifert revise the optimal control problem and consider what corresponds to a generalization of a dynamical Schr\"odinger bridge \cite{DaiP1991,LeoC2014}.

Nevertheless, the idea that optimal control of the mean work in stochastic thermodynamics leads to unphysical or non-operational solutions continues to be found in the contemporary literature. This idea contradicts the investigation and control of small system processes, which requires high-precision experimental setups. If a mathematical description of a physical process yields unphysical results, some necessary elements must be missing in the model. 

Why, then, are solutions to the mean work unphysical, while 
solutions to the mean heat minimisation problem between prescribed states are not? 

The purpose of this note is to answer this question in a conclusive way. The main conceptual takeaway is the following.
The mean work depends upon the variation of the internal energy of the system. Within stochastic thermodynamics, we recover this established fact using stochastic calculus, which requires functions to be at least once continuously differentiable in time and twice in space. Under these hypotheses, following Sekimoto \cite{SekK1998}, we can write the first law at the level of stochastic trajectories and, upon averaging over realizations, in mean value sense. 

If we now regard the work mean value as a cost function, the variation of the internal energy constitutes a boundary cost. This implies that we should interpret tunable parameters in the internal energy as controlled state variables and not directly as controls. There are compelling mathematical and physical reasons for this viewpoint. Mathematically, we obtain a standard Bolza structure of the optimal control problem \cite[\S~3.3.2]{LibD2012}. This is the natural setting for applying Pontryagin's maximum principle \cite[\S~4.2]{LibD2012}, or dynamic programming methods \cite[\S~5]{LibD2012}. 

Physically, continuous differentiability in time of the driven internal energy is a minimum requirement to impose an actuator rate saturation constraint (a constraint on how fast the actuator, the physical hardware imposing the control on the system, can change). This is a  well-known condition in applied optimal control theory \cite[\S~6]{SaStSa2012}. We refer to this requirement as imposing speed limits. The diffusion approximation underlying overdamped and underdamped stochastic thermodynamics relies on a separation between the externally imposed protocol, the resolved system dynamics, and the unresolved degrees of freedom of the thermal baths represented by the Wiener process. Within this time-scale separation, it is justified to consider transitions that are orders of magnitude faster than the open system relaxation time. This is the case for swift engineered equilibration recently experimentally demonstrated \cite{MaPeGuTrCi2016,GuJaPlPrTr2023}. Conversely, it is not justified to synthesize controls by means of infinite-bandwidth endpoint operations. An instantaneous quench may
be useful in fluctuation-relation arguments \cite[\S~4.7]{PePi2020}, but in fastest-control regimes operational protocols must be resolved as finite-time actuator motions. 

The physical importance of speed limits is supported by experimental evidence. A large body of evidence in quantum control demonstrates that trying to approximate controls with unbounded speed results in unwanted signal distortion, see, e.g., \cite{KoSkKhGlLu2008,BoCo2012,PaPu2014,HiCa2018} and \cite[\S~3.7.4]{DAlD2008}. The importance of time differentiability for operational fast protocols in stochastic thermodynamics is emphasized in \cite{GuJaPlPrTr2023}. In macrosystem optimal control, ignoring actuator rate saturation constraints leads to catastrophic consequences \cite{KaHa1998}. These considerations motivate treating the admissible class of controls as part of the physical model. In particular, actuator rate constraints are not merely numerical regularizations; they encode operational limitations of experimentally realizable protocols. 

It is fair to add that the notion of speed limits we are using here is broader than the one used in \cite{ShFuSa2018,VaVuSa2023}. There, speed limits
are conceptualized as a bound on the time it takes to drive an initial state into a target state obtained in terms of the mean entropy production. Such a bound can be regarded as a stochastic thermodynamic counterpart of time-energy uncertainty relations in quantum mechanics, see, e.g., \cite{PfFr1995}.

A further motivation for our work comes from significant developments in computational methods, including machine learning and, in particular, evolutionary reinforcement learning, see, e.g., \cite{BaWhCiBe2025} and \cite{LyRaCr2025}. In realistic setups, these methods make it possible to compute useful heuristic, when not optimal, control protocols for work extraction from small systems. However, development of numerical methods may be difficult in physically interpretable cases with two-time boundary conditions. For this purpose, mathematically well-posed problems, free of interpretation ambiguities and unnecessary numerical stiffness, are needed for benchmarking and testing. In this case, it may be interesting to look at problems without physical interpretation, that nevertheless share similar properties. 
To make our case clearly, we focus our analysis on the simplest possible case considered in \cite{ScSe2007}: a particle in a moving Gaussian trap in the overdamped regime. We refer to \cite{SaMG2025b} for an extension of our analysis to a case of underdamped dynamics, a problem introduced in \cite{GoScSe2008}, that has direct interest for experimental applications like \cite{MaPeGuTrCi2016,RaGuGuOdTr2023}.

\section{The work functional.} We consider a particle in one dimension driven by a mechanical force in the overdamped approximation 
\begin{align}
	\mathrm{d}\mathscr{q}_{t}=-\partial_{\mathscr{q}_{t}} U_{t}(\mathscr{q}_{t})\,\mathrm{d} t +\mathrm{d}\mathscr{w}_{t}
	\label{sde}
\end{align}
where $\mathscr{w}_{t}$ is a Wiener process \cite{PePi2020}. For the sake of simplicity, all quantities are non-dimensional.
 
Following \cite{PePi2020}, the average work done on the particle in a time interval $[0,\tf]$ is 
\begin{align}
	\mathcal{W}_{\tf}=\operatorname{E}\left[\int_{0}^{\tf}\!\mathrm{d}t\,\partial_{t}U_{t}(\mathscr{q}_{t})\right]
\end{align}
The partial derivative with respect to time $t$ only acts on the explicit time dependence of the mechanical potential $U_{t}$, which we always express by means of a subscript. For a particle in a Gaussian trap, the potential is simply
\begin{align}
	U_{t}(x)=\frac{(x-\lambda_{t})^{2}}{2}
	\label{trap}
\end{align}
The time dependent function $\lambda_{t}$ denotes the centre of the trap. In such a case, upon denoting 
\begin{align}
	\mathscr{x}_{t}=\operatorname{E}\left[\mathscr{q}_{t}\right]
	\label{mean}
\end{align}
  as the mean value of trajectories of (\ref{sde}) subject to (\ref{trap}), the work functional reduces to
  \begin{align}
  	\mathcal{W}_{\tf}=\int_{0}^{\tf}\!\mathrm{d}t\, \dot{\lambda}_{t}\,\left(\lambda_{t}-\mathscr{x}_{t}\right)
  \label{work}
  \end{align}
 As discussed in the introduction, the above derivation of the work holds for the case of a particle in a moving laser trap discussed in \cite{ScSe2007}. We refer readers unfamiliar with this idea to \ref{app:Gaussian} for further details on the derivation of the expression of the mean work.

\section{The {\textquotedblleft}naive{\textquotedblright} formulation of the work optimal control problem.} 
\label{sec:naive}

A very natural question is how to steer a thermodynamic transition by performing the minimum amount of mean work on the system. If we identify (\ref{work}) as a cost function, we are confronted with two issues: the first is assigning boundary conditions specifying initial and target system state, and the second is identifying the class of admissible controls that we can use to guide the transition. Physically, we may have in mind a situation where the system is in an equilibrium state at an initial time $t=0$. This means that initially the drift in (\ref{sde}) is proportional to the gradient of the logarithm of the initial probability distribution (its {\textquotedblleft}score function{\textquotedblright}). In other words, the initial current velocity of the system \cite[\S~13]{NelE2001} should vanish. The final state can be out of equilibrium, instead determined by physical requirements on the target state of the system and work minimisation. 

The coexistence of these two, very plausible, requirements for the boundary conditions is intertwined with the question about the set of admissible controls. This is true in particular when considering that (\ref{work}) depends on both the trap centre $\lambda_{t}$, and its time derivative $\dot{\lambda}_{t}$. Which of these is the physically relevant control of a transition at minimum work?

To answer this question, we may try to rewrite the work in terms of a running cost, specified by the mean released heat,
\begin{align}
	\mathscr{Q}_{\tf}=\int_{0}^{\tf}\!\mathrm{d}t\,
	(\mathscr{x}_{t}-\lambda_{t})^{2}
	\label{heat}
\end{align}
 and a terminal cost, the change of mean internal energy:
 \begin{align}
 	\operatorname{E}\left[U_{\tf}(\mathscr{q}_{\tf})-U_{0}(\mathscr{q}_{0})\right]=\frac{(\mathscr{x}_{\tf}-\lambda_{\tf})^{2}-(\mathscr{x}_{0}-\lambda_{0})^{2}}{2}
 \end{align}
 We get
\begin{align}
	\mathcal{W}_{\tf}=\frac{(\mathscr{x}_{\tf}-\lambda_{\tf})^{2}-(\mathscr{x}_{0}-\lambda_{0})^{2}}{2}+\mathscr{Q}_{\tf}
	\label{work2}
\end{align}
 
Did we gain anything with the rewriting? From the optimal control theory point of view, the answer is clearly \emph{no}. The canonical form of optimal control theory handles problems with running and terminal cost in \emph{Bolza form} \cite{BolO1907}, see, e.g., \cite[\S~3]{LibD2012} for a modern presentation adapted to the case of interest. Bolza form requires that the terminal cost is only a function of system state variables. In this case, the only state variable is the mean value (\ref{mean}) of the system coordinate in configuration space. The terminal cost in (\ref{work2}) does not satisfy this requirement if we identify $\lambda_{t}$ as the control variable. 

We can try to proceed by imposing an initial condition on the system state variable
\begin{align}
	\mathscr{x}_{0}=0
	\label{naive:xinit}
\end{align}
and, ignoring the Bolza requirement, time-boundary conditions on the trap centre:
\begin{align}
	\lambda_{0}=0
	\label{naive:Linitial}
\end{align}
and
\begin{align}
	\lambda_{\tf}=\ell
	\label{naive:Lfinal}
\end{align}
We thus derive the cost functional
\begin{align}
		\widetilde{\mathcal{W}}_{\tf}=\frac{(\mathscr{x}_{\tf}-\ell)^{2}}{2}+\int_{0}^{\tf}\!\mathrm{d}t\,
	(\mathscr{x}_{t}-\lambda_{t})^{2}
	\label{naive:htc}
\end{align}
 By working with (\ref{naive:htc}), we can apply standard optimal control theory, for instance using Pontryagin's principle. As the problem does not give rise to abnormal extremals, see, e.g. \cite{BaBePlRaTr2025}, applying Pontryagin's principle means finding extremals of the action functional
\begin{align}
	\mathcal{A}_{\tf}=\frac{(\mathscr{x}_{\tf}-\ell)^{2}}{2}+\int_{0}^{\tf}\!\mathrm{d}t\,
	\Big{(}(\mathscr{x}_{t}-\lambda_{t})^{2}+\mathscr{y}_{t}\left(\dot{\mathscr{x}}_{t}-\lambda_{t}+\mathscr{x}_{t}\right)\Big{)}
\end{align}
As customary, we refer to the Lagrange multiplier $\mathscr{y}_{t}$ in Pontryagin's action as the co-state variable. The first order conditions for state and co-state variables (see e.g. \cite[\S~4]{LibD2012}) are
\begin{align}
&	\dot{\mathscr{x}}_{t}=\lambda_{t}-\mathscr{x}_{t} && \dot{\mathscr{y}}_{t}=\mathscr{y}_{t}-2\,( \lambda_{t}-\mathscr{x}_{t})
	\label{naiveqs}
\end{align}
Trivially, the equation for the state variable $\mathscr{x}_{t}$ recovers the equation for the mean value \eqref{mean} of the Gaussian process \eqref{Gaussian:sde}.
 The equations (\ref{naiveqs}) are complemented by the stationary condition
\begin{align}
	\mathscr{y}_{t}=2\,(\lambda_{t}-\mathscr{x}_{t})
	\label{naivestat}
\end{align}
and, more importantly for the present discussion, 
 the boundary condition
\begin{align}
	\mathscr{x}_{\tf}-\ell=-\mathscr{y}_{\tf}
	\label{naivebc}
\end{align}
The minimisation of \eqref{naive:htc} satisfying~\eqref{naivebc} is a mathematically well-posed problem. We clearly see that $\lambda_{\tf}$ does not enter the boundary condition. A reader unfamiliar with our initial discussion would therefore have no reason to relate $\ell$ to $\lambda_{\tf}$. Namely, upon inserting the stationary condition (\ref{naivestat}) into (\ref{naiveqs}), we find
\begin{align}
&	\dot{\mathscr{y}}_{t}=0 && \forall t\in[0,\tf]
\end{align}
 and arrive at the explicit unique solution 
\begin{align}
	\begin{split}
&		\mathscr{x}_{t}=\frac{\ell\,t}{\tf+2} 
		\\
&		\lambda_{t}=\ell\,\frac{t+1}{\tf+2}
	\end{split}
	&& \forall t\in[0,\tf]
	\label{naivesol}
\end{align}
of the two-time boundary problem. For the cost functional (\ref{naive:htc}), this evolution yields
\begin{align}
	\widetilde{\mathcal{W}}_{\tf}=\frac{\ell^{2}}{\tf+2}
	\label{worknaive}
\end{align}
which is the result found in \cite{ScSe2007}. From this calculation, we can draw some conclusions 
 that can apply both to general overdamped or underdamped dynamics.
\begin{enumerate}[style=unboxed,leftmargin=0cm,label={\upshape\bfseries L-\roman*}]
	\item \label{L-Bolza} The terminal cost imposes conditions on state and co-state variables at the end of the horizon. The reason why the Bolza form of optimal control \cite[\S~3]{LibD2012} requires the terminal cost to be independent of controls is thus evident. Variations of the action with respect to controls specify the relation that these need to satisfy with state and co-state variables in the control horizon $[0\,,\tf]$. Violating the Bolza structure requirements leads to an additional condition on the controls that generically makes the problem ill-posed from the point of view of standard optimal control theory. Bolza form is necessary to prove the equivalence between optimal control problems with terminal and running cost and optimal control problems in Mayer form, where only a terminal cost appears \cite[\S~3.3.2]{LibD2012}. On its turn, Mayer form is the starting point to prove Pontryagin's maximum principle \cite[\S~4.2]{LibD2012}: the principle cannot be invoked if Bolza structure is violated. In other words: in standard control theory jumps of the control produce extremals consisting of distinct arcs. Gluing arcs is subject to the constraints imposed by the well-known Weierstrass-Erdmann conditions (continuity of co-state and Pontryagin pre-Hamiltonian at switching time) \cite[\S~3.1.1]{LibD2012}. In our case, Weierstrass-Erdmann require $\mathscr{y}_{\tf}$ to vanish which is incompatible with \eqref{naivebc}. The upshot is that the infinite speed arc required to salvage the work interpretation is inconsistent with standard optimal control theory.
  
	\item \label{L-inf} Let's take the perspective that the observations in \ref{L-Bolza} are just 
  technicalities. Let's also ignore that any control actuator is rate limited and that, as discussed in the introduction, there is an ample literature documenting the detrimental consequences of trying to unconditionally approximate infinite bandwidth control. We instead suppose that we can perform an instantaneous quench by deforming (\ref{naivesol}) on a set of zero Lebesgue measure at both ends of the control horizon. By doing so, we include control jumps at $t=0$ to ensure consistency with equilibrium at $t=0$, i.e., eqs. (\ref{naive:xinit}), (\ref{naive:Linitial}), and at $t=\tf$, such that $\lambda_{\tf}=\ell$, i.e, (\ref{naive:Lfinal}). 
	In fact, a mathematically minded reader may describe what we are doing as taking the infimum over protocols $\lambda_{t}$ that satisfy the conditions  (\ref{naive:Linitial}) and (\ref{naive:Lfinal}). Because of the existence on the \emph{closed} interval $[0,\tf]$ of the mathematically well-posed minimisation problem~\eqref{naive:htc} satisfying (\ref{naivebc}), it is clear that the search for the infimum must converge to (\ref{naivesol}) and (\ref{worknaive}) in the \emph{open} interval $(0,\tf)$. 
	\item \label{L-ambiguous} If we accept the reasoning in \ref{L-inf}, then, as soon as we turn to physical applications, we have to face the problem that Schmiedl and Seifert encountered in \cite{ScSe2008}. The optimal protocol includes an unphysical branch (protocols varying at infinite speed) which yields a finite contribution to the internal energy and directly affects the first law. As a consequence, the computation of physical quantities, for example, the efficiency at maximum power for a Carnot cycle, is ambiguous. 
  In \cite{ScSe2008}, Schmiedl and Seifert's solution to these ambiguities is clear: they abandon the work problem, and instead consider the minimisation of the heat released (\ref{heat}), by imposing two-time boundary conditions on the state variable (\ref{mean}).
\end{enumerate}
We have encountered the following chain of implications:
\begin{align*}
	&\mbox{Violation of standard Bolza structure}
  \\&\hspace{3.0cm}
  \Longrightarrow\mbox{Endpoint operations at infinite speed}
  \\&\hspace{6.0cm}
  \Longrightarrow\mbox{Thermodynamic ambiguities}
	\nonumber
\end{align*}
In other words, the argument in \ref{L-inf} is mathematically acceptable but in practice of very little physical utility. 
 
Conceptual problems remain. On the one hand, in experimental applications, it is often natural to prepare the initial state of the system to be in equilibrium. On the other, it is physically unsatisfactory to give up the idea of considering transitions at minimum work. In our view, the above theoretical description does not take into account a physical requirement that is obvious to experimentalists: control protocols are subject to speed limits. We discuss this further in Section~\ref{sec:sl}.

\subsection{The Schrödinger bridge formulation.}

Let us briefly recall the minimisation of the mean heat functional (\ref{heat}) applied in \cite{ScSe2008}. First we observe that (\ref{heat}) specifies a positive definite cost, convex in the control variable specified by the trap centre. We impose two-time boundary conditions on the system state variable
\begin{align}
	 &\mathscr{x}_{0}=0, \qquad&\mathscr{x}_{\tf}=x_{\mathscr{f}}
	\label{naive:xbc}
\end{align} 
The resulting optimisation problem is a generalization of the dynamical Schr\"odinger bridge problem \cite{LeoC2014}, because of the way we assign the running cost. The optimal protocol is
\begin{align}
	\begin{split}
		&	\mathscr{x}_{t}=\frac{t}{\tf}x_{\mathscr{f}}
		\\
		& \lambda_{t}=\frac{t+1}{\tf}x_{\mathscr{f}}
	\end{split}
	 && \forall t\in[0,\tf]
	\label{bridgesol}
\end{align} 
The evaluation of the work is now unambiguous: the terminal value of the internal energy is fully specified by (\ref{bridgesol}) without any discontinuity. We emphasize that the optimum transition does not specify a transition between equilibrium states: the conditions
\begin{align}
&\lambda_{0}=\mathscr{x}_{0} &&\lambda_{\tf}=\mathscr{x}_{\tf}
\label{naive:cv}
\end{align}
do not hold. But the conditions \eqref{naive:cv} are neither physically needed to interpret the mean heat functional nor used at any stage in the mathematical treatment of the problem! 

On a more conceptual level, we may regard the minimisation of the mean heat~\eqref{heat}, as a generalization of the idea of dynamical Schr\"odinger bridge. A dynamical Schrödinger bridge problem, whose ``static'' prototype was introduced by Schr\"odinger in his remarkable paper \cite{SchE1931} (English translation in \cite{ChMGSc2021}), involves the minimisation of a cost functional exclusively specified by a running cost convex in the control and subject to two-time boundary conditions on the state variable which in the infinite dimensional case coincides with the system probability density \cite{LeoC2014}. Upon generalizing the assignment of the running cost with respect to the reference case \cite{DaiP1991}, we can use the idea of dynamical bridge optimisation to determine lower bounds on the cost of a finite-time thermodynamic transition. Examples of applications are efficiency at maximum power \cite{ScSe2008}, and the study of Landauer's bound on the cost of the erasure of one bit of information \cite{AuGaMeMoMG2012}. In this sense, Schr\"odinger's paper \cite{SchE1931} should be considered the origin and a source of inspiration for optimal control in statistical physics.

\section{A refined formulation: inclusion of speed limits on controls.}
\label{sec:sl} 

We now suppose that we can vary the trap centre by only controlling the velocity $\upsilon_{t}$ with which it moves
\begin{align}
	\dot{\lambda}_{t}=\upsilon_{t}
	\label{speed}
\end{align}
A natural physical constraint is that the speed is bounded:
\begin{align}
&	|\upsilon_{t}|\,\leq\,V && \forall\,t
	\label{hard}
\end{align}
The existence of speed limits can be easily justified without invoking relativity. Namely, overdamped and underdamped dynamics are derived from an underlying Hamiltonian dynamics by means of multiscale perturbation theory \cite[\S~8]{PavG2014}. The derivation holds under the assumption of a strong separation between the time scales of the system dynamics and those of the environment modelled by the Wiener process in (\ref{sde}). 
Based on these considerations, we want $V$ to be finite to ensure time-scale separation from the environment, but sufficiently large to guarantee that boundary conditions are in the reachable set of the control problem \cite[\S~4.4.2]{LibD2012}. More explicitly, we require the heuristic bound
\begin{align}
	\infty\,>\,V\,\tf \,\gg\, \mbox{max}\left(|\mathscr{x}_{\tf}-\mathscr{x}_{0}|,|\lambda_{\tf}-\lambda_{0}|\right)
\end{align}
In the presence of speed limits, the mean value of the position of the system state variable $\mathscr{x}_{t}$ and the centre of the trap $\lambda_{t}$ are both \emph{state variables} of the control problem. We refer to them as the system and the parameter state variables, respectively. The internal energy is now a state function both from the thermodynamic point of view (by definition) and from the control point of view (because it depends only upon state variables and not upon controls). The action functional in the presence of speed limits now specifies a well-posed Bolza form problem
\begin{equation}\begin{split}
	\mathcal{A}_{\tf}^{\scriptscriptstyle{(s.\ell.)}}&=\frac{(\mathscr{x}_{\tf}-\lambda_{\tf})^{2}-(\mathscr{x}_{0}-\lambda_{0})^{2}}{2} \\
&	+\int_{0}^{\tf}\!\mathrm{d}t\,\Big{(}(\lambda_{t}-\mathscr{x}_{t})^{2}
	+ \mathscr{y}_{t}^{\scriptscriptstyle{(1)}}(\dot{\mathscr{x}}_{t}+\mathscr{x}_{t}-\lambda_{t})+\mathscr{y}_{t}^{\scriptscriptstyle{(2)}}(\dot{\lambda}_{t}-\upsilon_{t})\Big{)}\end{split}
	\label{sl:action}
\end{equation}
once complemented by the hard wall condition (\ref{hard}) on the control. To fully specify the optimisation problem, we need to impose boundary conditions on the extended set of four state variables.

Imposing the two-time boundary conditions on the system state variable (\ref{naive:xbc}) is certainly physically meaningful. In addition, we require (\ref{naive:Linitial}) to impose that the system starts at equilibrium. This may not necessarily be the case for all applications but is of sufficient general interest. We thus arrive at our central question: what are the boundary conditions on $\lambda_{\tf}$ that allow us to distinguish transitions between equilibrium states from transitions at minimum work?

\subsection{Optimal swift engineered equilibration}
\label{sec:see}

The case of a transition between equilibrium states is clear and was already discussed in \cite{AuMeMG2012}. In our example, the final condition describes a system in equilibrium with an environment at the same temperature at the beginning and 
end of the transition
\begin{align}
	\lambda_{\tf}=\mathscr{x}_{\tf}=x_{\mathscr{f}}
	\label{sl:equilibrium}
\end{align}
The set of four boundary conditions (\ref{naive:xinit}), (\ref{naive:Linitial}) 
and (\ref{sl:equilibrium}) specifies a very stylised model of optimal swift engineered equilibration \cite{MaPeGuTrCi2016}. Transitions between equilibria in finite time are certainly of great physical interest \cite{GuJaPlPrTr2023}. We do not, however, discuss this further here, because in the singular limit $V$ tending to infinity, it clearly recovers the well-posed Schr\"odinger bridge problem specified by (\ref{naive:xbc}).

\subsection{The minimum work problem (to reach a target state)}
\label{sec:mw}

Let us now assume boundary conditions (\ref{naive:xbc}) and (\ref{naive:Linitial}), but look for a transition that requires minimum work to be done to reach the target system final state. Correspondingly, the optimal value of (\ref{sl:action}) must be stationary with respect to a variation of $\lambda_{t}$. 
We obtain the condition
\begin{align}
	\lambda_{\tf}-\mathscr{x}_{\tf}+\mathscr{y}_{\tf}^{\scriptscriptstyle{(2)}}=0
	\label{realworkbc}
\end{align}
As always, the presence of a terminal cost imposes a boundary condition on a co-state, specifically $ \mathscr{y}_{t}^{\scriptscriptstyle{(2)}}$. Variations of (\ref{sl:action}) with respect to state and co-state variables yield four equations with four corresponding boundary conditions. We write these equations in \ref{app:synthesis}.

\subsection{Synthesis}

\begin{figure}[!h]
	\centering
	\includegraphics[width=0.9\linewidth]{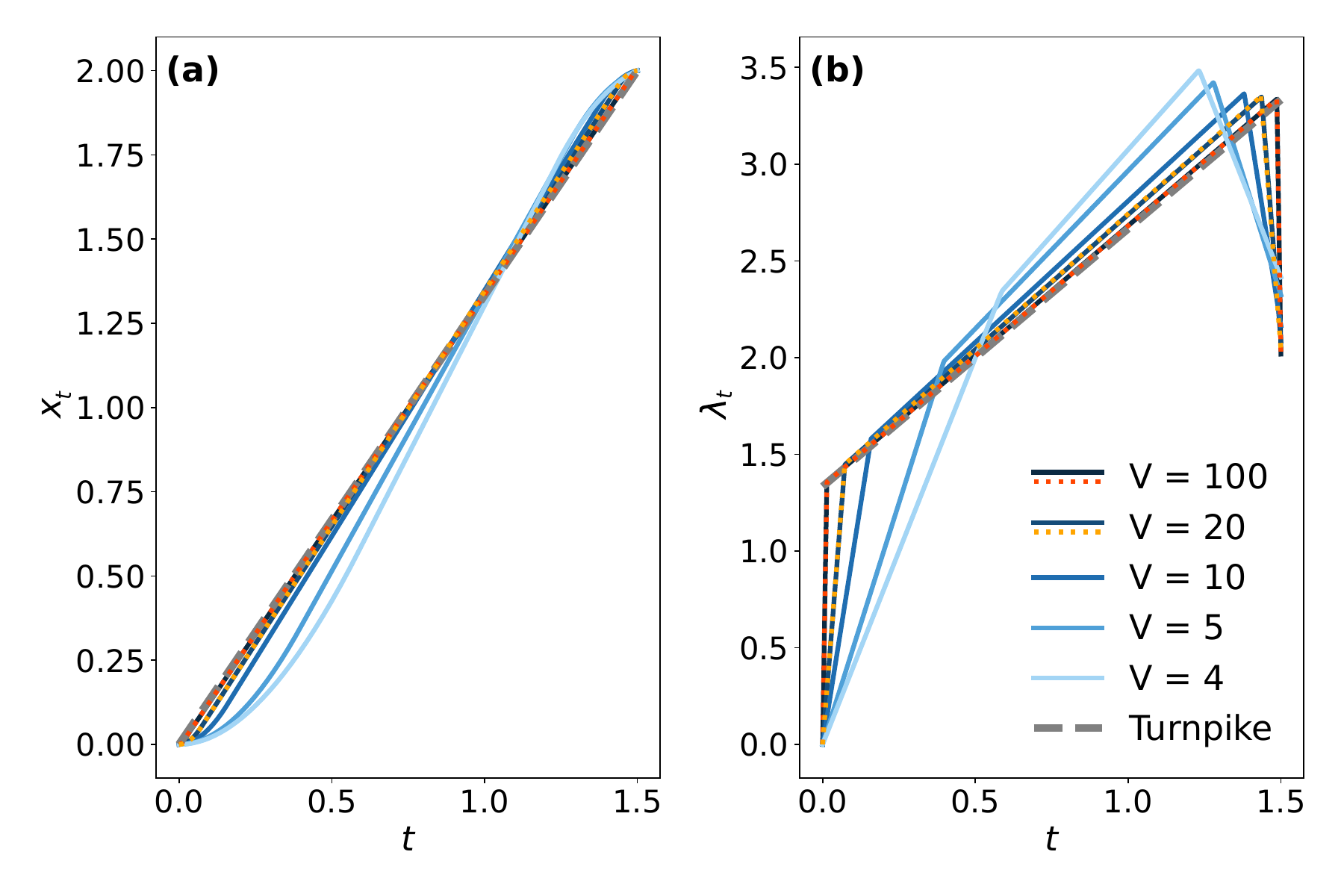}
	\caption{The mean value $\mathscr{x}_{t}$ (a) and trap centre $\lambda_{t}$ (b), determined by the boundary conditions (\ref{naive:Linitial}), (\ref{naive:xbc}) and (\ref{realworkbc}) with $x_{\mathscr{f}}=2$ and $\tf=1.5$. The blue curves are obtained by imposing speed limits on $\dot{\lambda}$ of the form~\eqref{hard} with increasing values for $V$, solved by a numeric direct optimisation method (see~\ref{direct_numerics}). We also show (orange dotted), the results of solving the corresponding transcendental equations arising from synthesis using perturbative expansion evaluated at $V=100$ (see~\ref{app:synthesis}). For larger values of $V$, the quantities stay closer to the turnpike solution (grey, dashed lines) specified by (\ref{bridgesol}) for longer during the control horizon. \label{Fig:hard}}
\end{figure}

The action (\ref{sl:action}) is not convex in the control $\upsilon_{t}$: we can conceptualize (\ref{hard}) as a hard wall potential confining controls in a compact interval.
The well-known consequence (see \cite{BoSiSu2021,HegG2013,BaBePlRaTr2025} and references therein) is that the optimal control strategy consists of gluing together {\textquotedblleft}push{\textquotedblright} regions where the trap centre speed saturates the bound (\ref{hard}), with {\textquotedblleft}turnpike{\textquotedblright} regions (see \cite{ZasA2014} for a justification of the terminology) where the stationary condition
\begin{align}
	\mathscr{y}_{t}^{\scriptscriptstyle{(2)}}=0
	\label{hardstat}
\end{align}
implicitly determines $\upsilon_{t}$ via the co-state $\mathscr{y}_{t}^{\scriptscriptstyle{(2)}}$. The process of optimally gluing extremals together is called synthesis \cite{BoSiSu2021}. In~\ref{app:synthesis}, we describe the synthesis procedure that leads to the optimal control strategies plotted in Fig.~\ref{Fig:hard}. 

In our example, linearity of the dynamics yields explicit expressions for extremals both in push and turnpike regions. The optimal times for when to glue the sections of the extremals are, however, determined by the roots of a transcendental equation. In~\ref{app:synthesis}, we also give the expression of the optimal control problem obtained by solving the transcendental equation in the limit of $V$ tending to infinity. This limit is singular because, in the absence of speed constraints, $\lambda_{t}$ changes its status and becomes a control. The remarkable consequence is that, in the limit $V$ tending to infinity, swift engineered equilibration and minimum work transition reduce to the same optimal control problem: the Schr\"odinger bridge solved by (\ref{bridgesol}). We discuss this point in further detail in Section~\ref{sec:hwm} below.

This definition of thermodynamic transition at minimum work is to us the most physically transparent. We nevertheless want to discuss some alternatives. 
The full specification of optimal protocols for the hard wall potential case (\ref{hard}) requires the solution of a transcendental equation, making the discussion somewhat cumbersome. Qualitative and, to a large extent, quantitative properties of an optimal control problem, including in this case the limit $V$ tending to infinity, are sensitive to the boundary conditions and not to whether the set of admissible controls 
is determined by a hard wall potential or by a softer confining potential. 
 
\subsection{Softening the hard wall potential on the controls}

For analytical results, it is convenient to {\textquotedblleft}soften{\textquotedblright} the hard wall potential (\ref{hard}) which confines controls to a compact set, with a harmonic potential. A harmonic potential is also an example of self-concordant barriers whose discovery has led to major developments in the design of fast algorithms of numerical optimisation \cite{NeTo2008}. 

We modify Pontryagin's action (\ref{sl:action}) to 
\begin{align}
	\tilde{\mathcal{A}}_{\tf}^{\scriptscriptstyle{(s.\ell.)}}=\mathcal{A}_{\tf}^{\scriptscriptstyle{(s.\ell.)}}+\int_{0}^{\tf}\!\mathrm{d}t\,\frac{\upsilon_{t}^{2}}{2\,V^{2}}
	\label{soft}
\end{align}
and we apply standard variational calculus to it. This allows us to replace (\ref{hardstat}), which only holds during turnpike intervals, with the stationary condition
\begin{align}
	\upsilon_{t}=V^{2}\,\mathscr{y}_{t}^{\scriptscriptstyle{(2)}}
\end{align}
which holds instead at any time during the control horizon. The first order conditions for optimisation are now specified by the analytically integrable linear Hamiltonian system
\begin{align}
\begin{split}
	& \dot{\mathscr{x}}_{t}=\lambda_{t}-\mathscr{x}_{t}
	\\
	& \dot{\lambda}_{t}=V^{2}\mathscr{y}_{t}^{\scriptscriptstyle{(2)}}
	\\
	&\dot{\mathscr{y}}_{t}^{\scriptscriptstyle{(1)}}=\mathscr{y}_{t}^{\scriptscriptstyle{(1)}}-2(\lambda_{t}-\mathscr{x}_{t})
	\\
	&\dot{\mathscr{y}}_{t}^{\scriptscriptstyle{(2)}}=-\mathscr{y}_{t}^{\scriptscriptstyle{(1)}}+2(\lambda_{t}-\mathscr{x}_{t})
\end{split}
\label{harmonic}
\end{align}
We now turn to the description of alternative optimal control problems that can be studied using the work as a cost.

\subsubsection{Transition at minimum work, again.}
\label{sec:hwm}

\begin{figure}[!h]
	\centering\includegraphics[width=0.9\linewidth]{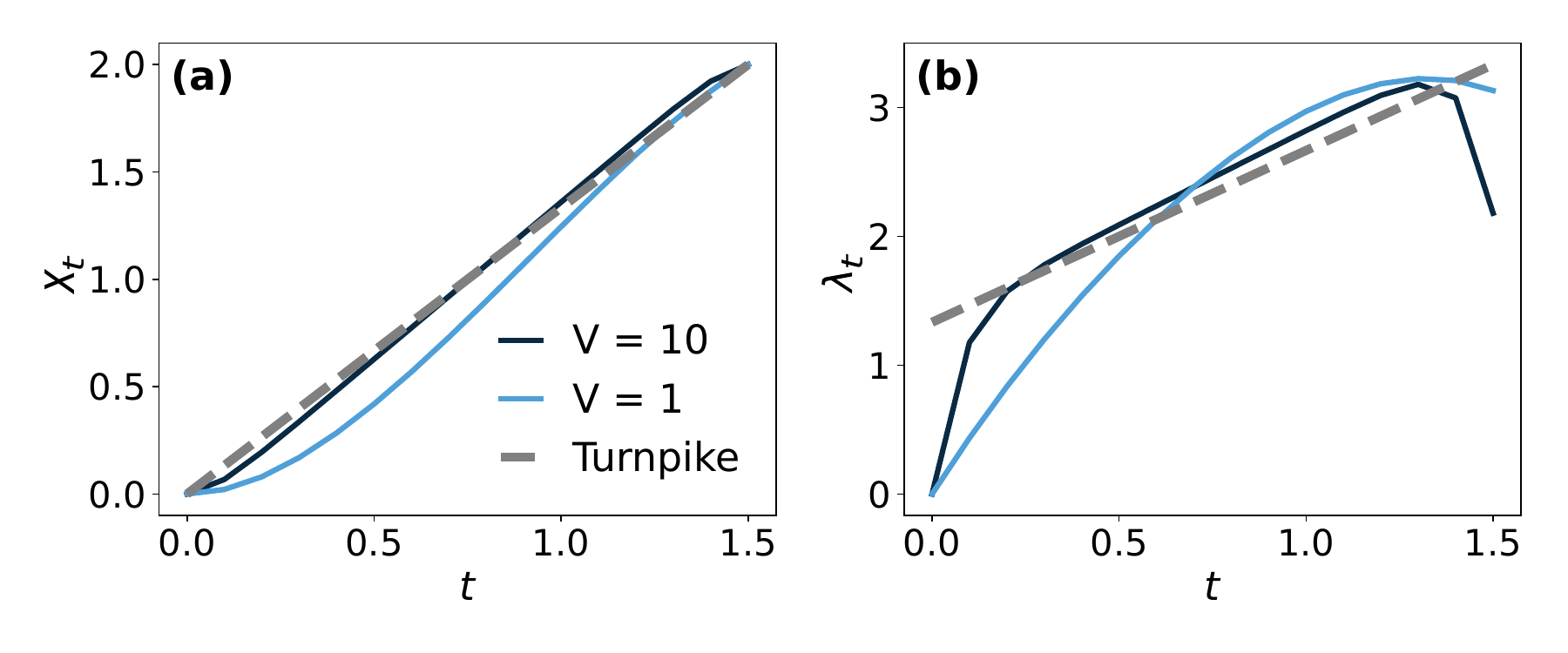}
	\caption{The mean value $\mathscr{x}_{t}$ (a) and trap centre $\lambda_{t}$ (b) determined by the boundary conditions (\ref{naive:Linitial}), (\ref{naive:xbc}) and (\ref{realworkbc}) with $x_{\mathscr{f}}=2$ and $\tf=1.5$. The curves are obtained by imposing a harmonic penalty with $V=1$ and $V=10$ as labelled. 
  As $V$ increases, the effect of distinct penalties on $\dot{\lambda}_{t}$ decreases, and $\mathscr{x}_{t}$ and $\lambda_{t}$ stay close to the {\textquotedblleft}turnpike{\textquotedblright} solution (grey, dashed lines), corresponding to (\ref{bridgesol}), for longer during the control horizon. Interestingly, as $V$ tends to infinity the limit value of $\lambda_{\tf}$ tends to $x_{\mathscr{f}}$ to approximate equilibrium. This is a feature of the model because the terminal cost is positive definite. \label{Fig:harmonic}}
\end{figure}

We return to the discussion on transitions at minimum work, now in the presence of the harmonic confining potential. Fig.~\ref{Fig:harmonic} shows the behaviour of solutions, qualitatively equivalent to those in Fig.~\ref{Fig:hard}. We give the exact explicit expressions of these solutions in \ref{app:hmw}. From these expressions, we recover
\begin{align}
	\begin{split}
		\lim_{V\uparrow \infty}	\mathscr{x}_{t}&=\frac{t}{\tf}x_{\mathscr{f}}	
		\\
		\lim_{V\uparrow \infty}	\lambda_{t}&=\frac{t+1}{\tf}x_{\mathscr{f}}	
	\end{split}
	&&
	\begin{split}
	 0\,\leq\,t\,\leq\,\tf
		\\
	0\,<\,t\,<\,\tf
	\end{split}
	\label{exact2}
\end{align}
The advantage of working with a soft penalty (\ref{soft}) is that we can now extract the properties of optimal protocols for finite $V$, qualitatively the same as and quantitatively close to those produced by the hard wall potential (\ref{hard}), from explicit analytic formulas. In agreement with the main findings of \cite{AuMeMG2012}, speed limits generate a boundary layer at both ends of the control horizon. The size of these boundary layers contracts exponentially as $V$ tends to infinity. The boundary layer ensures that all four boundary conditions (\ref{naive:xbc}), (\ref{naive:Linitial}), and (\ref{realworkbc}) are satisfied. The value of the work done to steer the system to a defined target state is equal to the mean heat release 
\begin{align}
	\lim_{V\uparrow\infty}\mathcal{W}_{\tf}=\lim_{V\uparrow\infty}\mathscr{Q}_{\tf}=\frac{x_{\mathscr{f}}^{2}}{\tf}
	\label{hwm:work}
\end{align}
The work done on the system coincides with the heat release because
\begin{align}
	\lim_{V\uparrow\infty}\frac{(\mathscr{x}_{\tf}-\lambda_{\tf})^{2}}{2}\equiv\lim_{V\uparrow\infty}\lim_{t\uparrow \tf }\frac{(\mathscr{x}_{t}-\lambda_{t})^{2}}{2}=\lim_{V\uparrow\infty}\frac{(x_{\mathscr{f}}-\lambda_{\tf})^{2}}{2}=0
\end{align}
As in the hard wall potential case, swift engineered equilibration and minimum work become the same in the singular limit of no speed limits. This justifies the physical intuition behind using the Schr\"odinger bridge problem to describe efficiency at maximum power in \cite{ScSe2008}. We remark, however, that we are assuming a linear drift in (\ref{sde}) to prove (\ref{hwm:work}).

Finally we notice that \eqref{hwm:work} is proportional to the action of a classical free particle of unit mass travelling from the origin to $x_{\mathscr{f}}$ in time $\tf$. Accordingly, it diverges if we take the limit $\tf$ tending to zero. Physically, this makes sense: an unphysical transition occurring at infinite speed has an infinite cost.

\subsubsection{Conditional work minimisation.}
\label{sec:hcwm}

\begin{figure}
  \centering
  \includegraphics[width=0.9\linewidth]{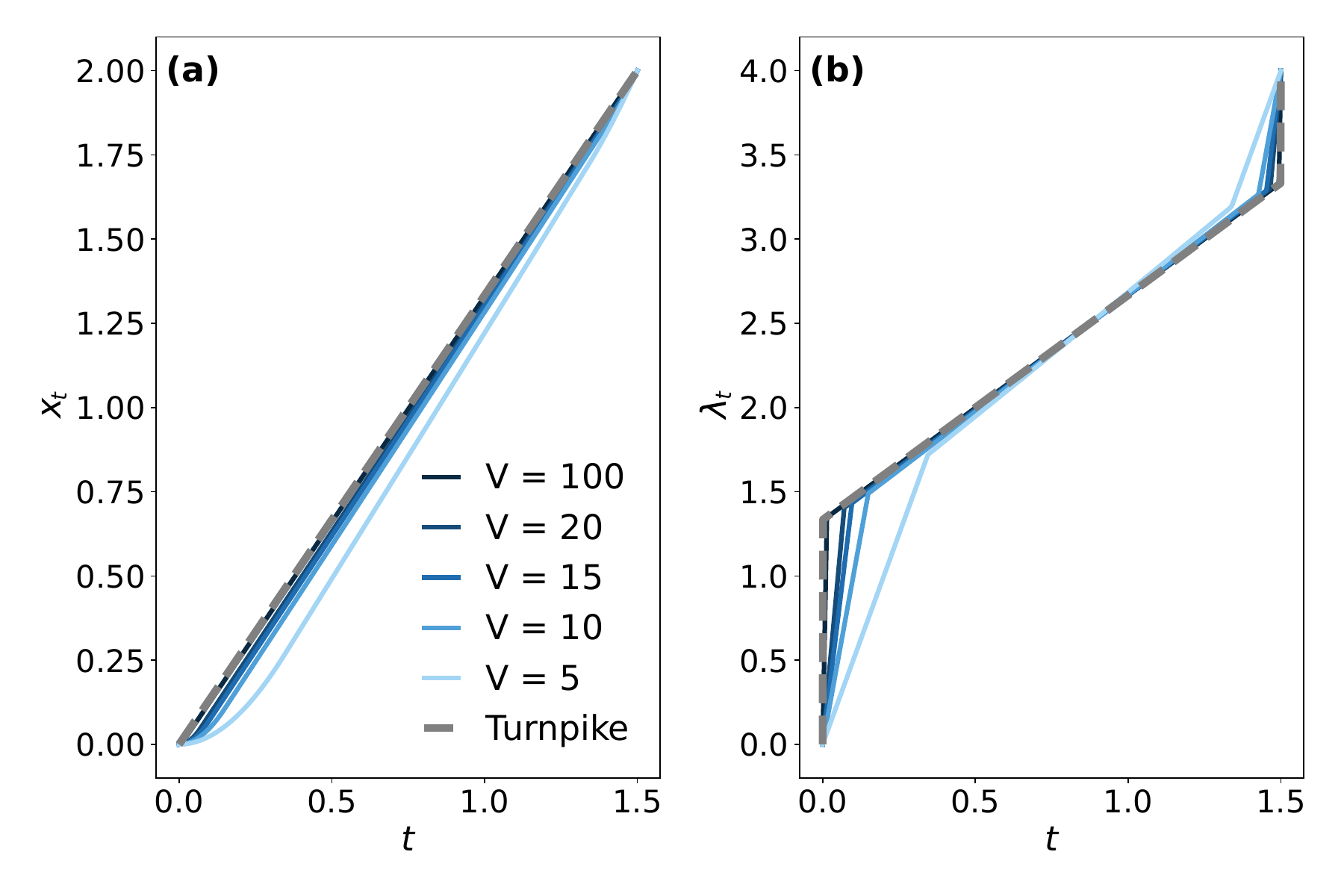}
  \caption{The mean value $\mathscr{x}_{t}$ (a) and trap centre $\lambda_{t}$ (b) determined by the boundary conditions (\ref{naive:xbc}), (\ref{naive:Linitial}), and (\ref{naive:Lfinal}) with $x_{\mathscr{f}}=2$, $\ell=4$ and $\tf=1.5$. Solutions for $V$ are computed with a hard penalty of the form~\eqref{hard} using a direct method. 
  As $V$ increases, $\mathscr{x}_{t}$ and $\lambda_{t}$ approach the solution of the first order conditions in the limit of no speed limits, corresponding to (\ref{bridgesol}).}
  \label{fig:condwork}
\end{figure}
Alternatively to minimising the work with respect to the final value of the parameter state variable, we can consider a transition which steers the parameter state variable towards a preassigned value. Mathematically, this amounts to imposing (\ref{naive:Lfinal}) instead of (\ref{realworkbc}).
The solution of the first order conditions (\ref{harmonic}) is amenable to the form
\begin{align}\label{eq:condworksol} \begin{split}
&	\mathscr{x}_{t}=x_{\mathscr{f}}\Big{(}f_{t}^{\scriptscriptstyle{(1)}}-f_{t}^{\scriptscriptstyle{(2)}}\Big{)}+\ell\,f_{t}^{\scriptscriptstyle{(2)}}
 \\
&	\lambda_{t}=x_{\mathscr{f}}\Big{(}f_{t}^{\scriptscriptstyle{(4)}}-f_{t}^{\scriptscriptstyle{(2)}}-f_{t}^{\scriptscriptstyle{(3)}}\Big{)}+\ell\, \Big{(}f_{t}^{\scriptscriptstyle{(1)}}+f_{t}^{\scriptscriptstyle{(2)}}+f_{t}^{\scriptscriptstyle{(3)}}\Big{)}\end{split}
\end{align}
We give the explicit expressions of the functions $f_{t}^{\scriptscriptstyle{(i)}}$, $i=1,\dots,4$ in \ref{app:hcmw}.
If we remove speed limits, we obtain
\begin{align}\begin{split}
  &	\lim_{V\uparrow\infty}	f_{t}^{\scriptscriptstyle{(1)}}=\frac{t}{\tf} \\
&\lim_{V\uparrow\infty}	f_{t}^{\scriptscriptstyle{(2)}}=0 
\end{split}
 && 0\,\leq\,t\,\leq \,\tf 
\end{align}
and
\begin{align}\begin{split}
&\lim_{V\uparrow\infty}f_{t}^{\scriptscriptstyle{(3)}}=-\frac{t}{\tf}
\\
&\lim_{V\uparrow\infty}	f_{t}^{\scriptscriptstyle{(4)}}=\frac{1}{\tf}
	\end{split}
  && 0\,<\,t\,<\,\tf 
\end{align}
We emphasize, however, that
\begin{align}
&\lim_{V\uparrow\infty}\lim_{t\downarrow0}f_{t}^{\scriptscriptstyle{(i)}}=\lim_{V\uparrow\infty}\lim_{t\uparrow\tf}f_{t}^{\scriptscriptstyle{(i)}}=0
&& i=3,4
\end{align}
Correspondingly, we find
\begin{align}
	\lim_{V\uparrow\infty}	\mathcal{W}_{\tf}=\frac{(x_{\mathscr{f}}-\ell)^{2}}{2}+\lim_{V\uparrow\infty}	\mathscr{Q}_{\tf}=\frac{(x_{\mathscr{f}}-\ell)^{2}}{2}+\frac{x_{\mathscr{f}}^{2}}{\tf}
	\label{hcwm:work}
\end{align}
The difference from the full work minimisation in Section~\ref{sec:hwm} is that the terminal cost is now fully specified by the boundary conditions and it is therefore insensitive to the limit. As expected, the work done on the system is greater than for the boundary condition (\ref{realworkbc}). 

The limit (\ref{hcwm:work}) assigns a value to the work done during a transition which, in the absence of a speed limit, is not well-defined. The fact that (\ref{hcwm:work}) recovers the minimum value (\ref{hwm:work}) when $\ell$ is equal to $x_{\mathscr{f}}$ does not imply that (\ref{hwm:work}) describes a transition between equilibria. The Schr\"odinger bridge problem is well-posed without any reference to equilibrium. Instead, we interpret that (\ref{hwm:work}) gives a lower bound on work and heat. 

\subsubsection{Work minimisation over the system final state for fixed terminal state of the actuator.}

In the presence of speed limits, $\lambda_{t}$ is a (parameter) state variable. Hence, we can also ask what is the value of the work if we impose the initial boundary conditions (\ref{naive:xinit}), (\ref{naive:Linitial}) and a final boundary condition on the trap centre (\ref{naive:Lfinal}) and optimize with respect to the system state variable. In this case, the fourth boundary condition is
\begin{align}
	\mathscr{x}_{\tf}-\ell+\mathscr{y}_{\tf}^{\scriptscriptstyle{(1)}}=0
	\label{sl:wmsbc}
\end{align}
We write the solution of equations (\ref{harmonic}) under these boundary conditions in \ref{app:wmsv}. Upon removing speed limits we now find
\begin{align}
	&	\lim_{V\uparrow \infty}	\mathscr{x}_{t}=\frac{\ell \,t}{\tf+2}
	&& 
		0\,\leq\,t\,\leq\,\tf
\end{align}
and
\begin{align}
	&
		\lim_{V\uparrow \infty}	\lambda_{t}=\ell \frac{t+1}{\tf+2}
	&&
		0\,<\,t\,<\,\tf
\end{align}
This is the optimal protocol originally proposed in \cite{ScSe2007}. In addition, we find that
\begin{align}
	\lim_{V\uparrow\infty}\frac{(\mathscr{x}_{\tf}-\lambda_{\tf})^{2}}{2}=\lim_{V\uparrow\infty}\frac{(\mathscr{x}_{\tf}-\ell)^{2}}{2}=\frac{2\,\ell^{2}}{(\tf+2)^{2}}
\end{align}
which recovers (\ref{worknaive}): we thus provided an explicit construction formalizing the argument \ref{L-inf}.

While mathematically well-posed, this control problem is not adapted to describe transitions meant to produce a desired target variation in the state of the system (e.g., an isothermal expansion). Analogously: does it make sense to look for the least-effort strategy for getting water flow from a pipe based on the angle we intend to turn the tap, rather than on the amount of water that comes out, including no water flowing as an admissible result? Usually, the purpose of (optimal) control is to steer a system toward a desired state by tuning parameters of the dynamics. Minimizing the work over the system final state variable for a fixed value of the parameter state variable seems therefore to swap the roles of the means and ends of control. 
 
 This is the scenario literally described by \eqref{worknaive} for vanishing horizon when we interpret it as the limit $V$ tending to infinity: an arbitrarily fast protocol that requires finite work to produce no change in the state variable of the system. In this limit a possible physical interpretation emerges. Minimizing the work with respect to the terminal value of the state of the system with a fixed state of the control may be interpreted as a lower bound on the cost of a quench.
 
 \section{Speed limits versus constraints on the controls}
 \label{sec:constraints}

 We start by summarizing the main points of the above discussion. When the total cost of a transition is the sum of a terminal and a running cost:
 \begin{enumerate}[style=unboxed,leftmargin=0cm,label={\upshape\bfseries T-\roman*}]
\item \label{T:co-state} We derive two-time boundary conditions by varying the total cost with respect to the state variables. In this way, we generically obtain equations for the final and initial values of co-states.
\item \label{T:control} We determine the optimal controls on the entire horizon by considering extremal values of the running cost with respect to the controls.
\item \label{T:Bolza} The Bolza form of optimal control forbids the boundary cost to depend on controls. This avoids any conflict between the control terminal values as determined by \ref{T:control} and the two time boundary conditions derived according to \ref{T:co-state}.
\item In stochastic thermodynamics, boundary costs depend on the internal energy or similar indicators that have an exact differential. Taking into account speed limits ensures that time dependent parameters subject to the control in the terminal cost are state variables. In this way, we comply with \ref{T:Bolza}, and deal with a well-posed problem. Most importantly, considering speed limits enables us to unambiguously discriminate between mathematical models of physical processes describing transitions at minimum mean work and transitions between equilibria at minimum mean entropy production \cite{SaMG2025b}.
\item \label{T:math}
Introducing speed limits increases the number of differential equations, thus \emph{changing the differential order of the first order conditions} of the control problem.
 \end{enumerate}
 Speed limits can be regarded as a constraint on the set of admissible controls. 
 Generally in optimal control, and stochastic thermodynamics in particular, we may need to impose other types of constraints in order to satisfy certain physical requirements. For example, \cite{BaBePlRaTr2025} constrains the control, identified with the stiffness of a confining potential, to be always positive and bounded, so that it describes the constraints in a concrete experimental setup. Modelling experimental setups using constraints restricts the set of admissible controls when finding extremals as per \ref{T:control}. Constraints on the intensity of the controls neither contradict nor are they an alternative solution to the problems described in \ref{L-Bolza}--\ref{L-ambiguous}. 
 
 Any constraints confining controls to a compact set, just like speed limits given by hard wall potentials, naturally mean that extremals must be constructed using synthesis. Continuity of state and co-state variables can be obtained at the price of discontinuities in the time derivatives of controls. This produces, in some cases, bang-bang optimal control strategies \cite[\S~4.4]{LibD2012}. 
 
 Experimental evidence, albeit so far only in quantum control \cite{HiCa2018}, shows that trying to implement protocols with abrupt variations in the laboratory results in unwanted signal distortions. These distortions are likely due to the breakdown of the separation-of-scales hypotheses, under which models of open system dynamics are derived. Softening hard wall constraints into confining potential penalties means that we can address these questions \emph{within a control dynamics of the same differential order in time}. In the previous section, we used this to enforce speed limits. In \cite{SaMG2025b}, we study a model similar to that of \cite{BaBePlRaTr2025}, where we impose speed limits on the time derivative of the stiffness of a mechanical potential and at the same time confine the stiffness to a compact interval. When the control dynamics is non-linear, softening constraints is necessary: efficient numerical algorithms for direct optimisation model constraints with self-concordant potentials \cite{NeTo2008}.

 \section{Conclusions} 

Technological advances in nano-system control make it possible to study transitions such as swift engineered equilibration \cite{MaPeGuTrCi2016,GuJaPlPrTr2023} that can occur at the fastest possible scale permitted by control actuators. A concrete example may help clarify this point: the current control across a dc-SQUID in the underdamped phase diffusion regime experimentally demonstrated in \cite{KiNiClBuHePe2005}.
The characteristic time scale of the dynamics is set by the plasma frequency $\omega_{p}$. A control actuator is an arbitrary waveform generator (AWG) whose characteristic rise-time $t_{r}$ (the inverse of the saturation rate) is given in the specs provided by the manufacturer of the device. For state-of-the-art AWG, one verifies that $\omega_{p}t_{r}$ is order of unity.
The optimal control problem describing fast processes should therefore take into account saturation rate limitations, here referred to as speed limits. The importance of speed limits has been previously realized in other fields of applied optimal control, see, e.g., \cite{HiCa2018,KaHa1998,SaStSa2012}. Indeed, the existence of speed limits is a real physical phenomenon that affects stochastic thermodynamics \cite{ShFuSa2018,VaVuSa2023}. When the cost includes the difference between the thermodynamic state function evaluated at the initial and final instant of the control horizon, e.g., the system internal energy, we have a well-posed optimisation problem, provided that we comply with Bolza's formulation \cite[\S~3.3.2]{LibD2012} of optimal control. Concretely, this means that the state function must also be a pure function of the state variables of the control problem.

 As already noticed long ago \cite{ScSe2008}, ignoring this requirement leads to unphysical protocols and, in the case of internal energy, ambiguities in the evaluation of the first law. Taking speed limits into account allows us to precisely formulate optimal control models of experimentally realizable finite-time transitions, such as swift engineered equilibration \cite{MaPeGuTrCi2016,GuJaPlPrTr2023}. In \cite{SaMG2025b}, we show how including speed limits yields measurably different predictions for thermodynamic indicators of nano-oscillator underdamped dynamics for transitions between distinct system end-states at minimum work or those connecting equilibria. There, we focus on optimal transitions that involve controlling the stiffness of the mechanical potential rather than its centre as done here. In the underdamped context, this means solving eight coupled equations with as many time boundary conditions.

 The continuously improving resolution of measurements in stochastic thermodynamics, see, e.g., \cite{DaCiBe2023}, suggests that the predictions of \cite{SaMG2025b} may soon be experimentally tested.
 
A main takeaway of our analysis is that with or without speed limits, optimal control problems of transitions are well-posed as Schrödinger bridge problems. However, mathematically solving bridge problems, especially in the case of underdamped dynamics with non-Gaussian boundary conditions, is very challenging \cite{SaBaMG2024}. It is therefore useful to look for weaker formulations that may provide qualitatively similar information. Recently, \cite{PaTaTr2021} proposed the concept of a {\textquotedblleft}half-bridge{\textquotedblright}, where the terminal cost is specified by the Kullback-Leibler divergence between the probability distribution of system state variables generated by the dynamics at the end of the control horizon and a reference pre-assigned distribution. We may extend this concept to general thermodynamic transitions for any running cost with physical interpretation, for example, the mean entropy production. First order conditions are then complemented by an initial condition on the probability distribution of the system and a terminal condition on the value function, that in the infinite dimensional case generalizes the notion of co-state. Given the same running cost, these two-time boundary conditions are often numerically less challenging to implement than a Schrödinger bridge problem.

 In fact, a big challenge for many algorithms in solving half- and Schrödinger bridge problems is finding final conditions on the co-states, which usually is done by an iterative approach, see, e.g.,~\cite{CaHa2022}. Considering the minimisation of~\eqref{naive:htc}, or in fact any problem where the cost is specified by a running and a terminal cost in Bolza form, can provide a good starting point for developing numerical methods. This is a useful simplification to test algorithmic convergence, where a physical interpretation is unnecessary, and has applications in, e.g., generative models of machine learning \cite{HuLaVaEi2024}.

 \section{Acknowledgments}
 
 We are very grateful to Marco Baldovin for a very careful reading of the first version of this manuscript. His insightful comments motivated an extensive revision of the original version and, in particular, the inclusion of Section~\ref{sec:constraints}. We also gratefully acknowledge discussions with Luca Peliti, Dario Lucente and Jukka Pekola. 
 
 PM-G acknowledges support from CoE FiRST of the Research Council of Finland (funding decision number: 346305).

 JS is supported by a University of Helsinki funded doctoral researcher position, Doctoral Program in Mathematics and Statistics. 

 \emph{Use of AI:} We used Claude Opus 5.5 to check the final manuscript for typos, and no automatic edits were added. 
 
   \appendix
   
   \section*{Appendices}
 \setcounter{section}{0}

 \section{Gaussian dynamics}
 \label{app:Gaussian}

 The statistics generated by realizations of the stochastic differential equation
 \begin{align}
 	\mathrm{d}\mathscr{q}_{t}=-(\mathscr{q}_{t}-\lambda_{t})\,\mathrm{d} t +\sigma\,\mathrm{d}\mathscr{w}_{t}
 	\label{Gaussian:sde}
 \end{align}
 maps a Gaussian probability distribution of the process $\mathscr{q}$ at time $t$ equal zero into a Gaussian distribution at later time. Gaussian distributions are fully characterized by their variance 
 \begin{align}
 	\mathscr{v}_{t}=\operatorname{E}\left[\mathscr{q}_{t}^{2}\right]-(\operatorname{E}\left[\mathscr{q}_{t}\right])^{2}
 	\nonumber
 \end{align}
 and mean value defined by (\ref{mean}). A straightforward calculation shows that the 
 variance evolves according to
 \begin{align}
 	\dot{\mathscr{v}}_{t}=-2\,\mathscr{v}_{t}+\sigma^{2}
 \end{align}
 In order to derive the model of a particle in a moving laser trap introduced in \cite{ScSe2007} and considered in the main text, one assumes that the variance is at equilibrium:
 \begin{align}
& 	2\,\mathscr{v}_{\star}=\sigma^{2} && \forall\,t
 	\nonumber
 \end{align}
 In such a case
 \begin{align}
 	\operatorname{E}\left[(\mathscr{q}_{t}-\lambda_{t})^{2}\right]=\mathscr{v}_{\star}+(\mathscr{x}_{t}-\lambda_{t})^{2}
 \end{align}
 and, as a consequence, the variance does not appear in the expression of the mean work.

 \section{Brief description of synthesis}
 \label{app:synthesis}
 
 By first using a {\textquotedblleft}soft{\textquotedblright} penalty potential, we can usually get some intuition as to how to construct an optimal synthesis for {\textquotedblleft}hard{\textquotedblright} penalty counterparts.
 For $V$ sufficiently large, we thus expect synthesis to involve two outer {\textquotedblleft}push{\textquotedblright} regions and a bulk {\textquotedblleft}turnpike{\textquotedblright} region. The intuition behind the appearance of these distinct regions is the
 Weierstrass theorem, guaranteeing the existence of a global minimum of a continuous function whose domain is on a compact set.
 The minimum can then be a critical point or sit on the boundary of the compact set.
 
 We refer to \cite{BaBePlRaTr2025} for detailed discussions of synthesis in the context of underdamped stochastic thermodynamics and to \cite{HegG2013} for an earlier application to quantum control.
 
 \subsection{Push regions}
 
 Push regions correspond to controls taking values specified by the boundary of the compact set to which admissible protocols are restricted. In our case, the compact set is determined by the condition (\ref{hard}). The first order conditions are
 \begin{align}
 	\begin{split}
 		&	\dot{\mathscr{x}}_{t}+\mathscr{x}_{t}-\lambda_{t}=0
 		\\
 		& \dot{\lambda}_{t}-c\,V=0
 		\\
 		&	\dot{\mathscr{y}}_{t}^{\scriptscriptstyle{(1)}}+\dot{\mathscr{y}}_{t}^{\scriptscriptstyle{(2)}}=0
 		\\
 		& \dot{\mathscr{y}}_{t}^{\scriptscriptstyle{(2)}}+\mathscr{y}_{t}^{\scriptscriptstyle{(1)}}+2\,(\mathscr{x}_{t}-\lambda_{t})=0
 	\end{split}
 \end{align}
 with $c$ only taking values $\pm 1$. Upon differentiating the last equation, we can eliminate the first co-state from the system:  
 \begin{equation}\begin{split}
 	\ddot{\mathscr{y}}_{t}^{\scriptscriptstyle{(2)}}-\dot{\mathscr{y}}_{t}^{\scriptscriptstyle{(2)}}
 	+2\,(\lambda_{t}-\mathscr{x}_{t}-c V)=0
 	\end{split}
 \end{equation}
 We hypothesize that push regions are located at both ends of the control horizon.
 
 \subsection{Turnpike region} 
 
 Upon imposing (\ref{hardstat}), the first order conditions reduce to
 \begin{align}
 	\begin{split}
 		&	\dot{\mathscr{x}}_{t}+\mathscr{x}_{t}-\lambda_{t}=0
 		\\
 		& \dot{\lambda}_{t}-\upsilon_{t}=0
 		\\
 		&\mathscr{y}_{t}^{\scriptscriptstyle{(1)}}+2\,(\mathscr{x}_{t}-\lambda_{t})=0
 		\\
 		&\mathscr{y}_{t}^{\scriptscriptstyle{(2)}}=0
 	\end{split}
 \end{align}
 These equations allow us to determine $\upsilon_{t}$. A straightforward analysis yields
 \begin{align}\begin{split}
 &	\mathscr{x}_{t}=\tilde{x}_{\mathfrak{0}}+\tilde{v}\,t \\
 &\lambda_{t}=\tilde{\ell}_{0}+v+\tilde{v}\,t
\end{split} \end{align}
 where $ \tilde{x}_{\mathfrak{0}}$, $\tilde{v}$ and $v$ are some unknown constants. We fix these constants as well as those produced by the integration of the first order equations in the push regions by continuously gluing solutions together.
 
 We surmise that for $V$ finite but sufficiently large, the turnpike region occurs between $t_1$ and $\tf - t_2$ such that 
 \begin{align}
 	0\,\leq\,t_{1}\,\leq\,t\,\leq\,\tf-t_{2}\,\leq\,\tf
 \end{align}
 where $t_{1}$ and $t_{2}$ must be determined by synthesis.

 \subsection{Synthesis}
 
 We glue together the solutions in the three regions by imposing continuity of state and co-state variables. To generate the curve in Fig.~\ref{Fig:hard}, we use the solution on the two outer regions to impose the boundary conditions (\ref{naive:xbc}), (\ref{naive:Linitial}), and (\ref{realworkbc}). We are then left with solving transcendental equations for $t_{1}$, $t_{2}$ solvable for large $V$. These quantities admit an analytic expansion starting at 
 \begin{align}
 	\varepsilon=V^{-1}
 \end{align} 
In this limit, the work is minimal if $c=1$ in the outer region starting at $t=0$, and $c=-1$ for the outer region ending at $t=\tf$. For the minimum work specified by the boundary condition (\ref{realworkbc}) at leading order, we find
\begin{align}\begin{split}
	\mathscr{x}_{t}=\begin{cases}
	(e^{-t}-1+t)\,V	+O(V^{-2})&\hspace{0.5cm} 0\,\leq\,t\,\leq\,t_{1}
		\\[0.3cm]
		\dfrac{t}{\tf}x_{\mathscr{f}}+\dfrac{2\,t-\tf}{2\,\tf^{3}\,V}\,x_{\mathscr{f}}^{2}+O(V^{-2})&\hspace{0.5cm} t_{1}\leq\,t \,\leq\,\tf -t_{2}
		\\[0.3cm]
		(1+\tf-e^{\tf-t}-t)\,V	+x_{\mathscr{f}} +O(V^{-2})&\hspace{0.5cm} \tf -t_{2}\,\leq\,t \,\leq\,\tf
	\end{cases}\end{split}
\end{align}
and
\begin{align}\begin{split}
	\lambda_{t}=\begin{cases}
		t\,V	&\hspace{0.5cm} 0\,\leq\,t\,\leq\,t_{1}
		\\[0.3cm]
		\dfrac{1+t}{\tf}x_{\mathscr{f}}+\dfrac{2+2\,t-\tf}{2\,\tf^{3}\,V}\,x_{\mathscr{f}}^{2}+O(V^{-2})&\hspace{0.5cm} t_{1}\leq\,t \,\leq\,\tf -t_{2}
		\\[0.3cm]
		(\tf-t)\,V	+x_{\mathscr{f}} + O(V^{-1})&\hspace{0.5cm} \tf -t_{2}\,\leq\,t \,\leq\,\tf
	\end{cases}\end{split}
\end{align}
and
\begin{align}
&	t_{1}=\dfrac{x_{\mathscr{f}}}{V\,\tf}+\dfrac{2+\tf}{2\,V^{2}\,\tf^{3}}\,x_{\mathscr{f}}^{2}+O(V^{-3})
&& t_{2}=\dfrac{x_{\mathscr{f}}}{V\,\tf}+O(V^{-2})
\end{align}
When controls are confined in a very {\textquotedblleft}narrow{\textquotedblright} region, the optimal protocol can be a more complicated sequence of push and turnpike regions than the one explored here. This phenomenon has been encountered in, e.g., \cite{BaBePlRaTr2025}.

 \section{Numerical Methods for Optimal Control}\label{direct_numerics}
 Numerical methods for solving the optimal control problem described in this paper can be roughly divided into two categories: direct and indirect. Indirect methods mean solving the system of ordinary differential equations arising from the first order optimality conditions. Direct methods on the other hand refer to directly optimising a cost (e.g.,~\eqref{naive:htc}) over a discretised space. 
 
 An algorithm for solving the problem directly is Interior Point optimisation (IPOPT). By using a direct method, we can impose the hard penalty of the form~\eqref{hard} on the control without needing to perform synthesis manually. 

 Direct methods discretise the state $\mathscr{x}$ and the control $\upsilon$. IPOPT imposes hard constraints in the problem as smooth self-concordant barriers to the cost. The barrier function takes increasingly larger values as controls approach their maximum values, penalising the cost. This {\textquotedblleft}smoothed{\textquotedblright} problem then yields a set of Karush-Kuhn-Tucker conditions \cite[\S~5.5.3]{BoVa2004}, a set of equations that the optimal solution must satisfy. Solving these can be computationally expensive, as it requires calculating gradients over the discretised system. The Karush-Kuhn-Tucker system is then solved with, e.g., a Newton method to find a search direction. Parameters are updated towards this direction, and steps are repeated until convergence. 

 To compute the numerical examples with a hard penalty shown in Figs.~\ref{Fig:hard} and~\ref{fig:condwork}, and with a harmonic penalty shown in Fig.~\ref{Fig:harmonic}, we use the Julia library InfiniteOpt~\cite{PuZhHoZa2022} to parse the problem into IPOPT~\cite{WacA2009}. 
 
 \section{Explicit expression of the solutions in the harmonic case}
 
 We provide the explicit expressions of the functions discussed in the main text.
 In all the expressions, we use
 \begin{align}
 	\omega=\sqrt{2\,V^2+1}
 \end{align}
 The singular limit of no speed limit corresponds to $\omega$ tending to infinity.
 
 \subsection{Transition at minimum work}
 \label{app:hmw}
 
 If we impose the boundary conditions (\ref{naive:xbc}), (\ref{naive:Linitial}), and (\ref{realworkbc}), the expression of the state variables solving (\ref{harmonic}) is
 \begin{align}\begin{split}
 	\mathscr{x}_{t}=&x_{\mathscr{f}}\frac{ \omega\,t\,\sinh(\omega\,\tf)+\frac{2\,\omega}{\omega^{2}+1}\,\big{(}\sinh(\omega(\tf-t))-\sinh(\omega\tf)\big{)}}{\omega\left(\tf -\frac{2}{\omega^{2}+1}\right)\sinh(\omega\,\tf)+2\, \left(\frac{\omega^{2}\,\tf}{\omega^{2}+1}-1\right)\cosh(\omega\,\tf)+2} \\
 	&+	x_{\mathscr{f}}\frac{\cosh(\omega(\tf-t)) +1-\cosh(\omega t)+\left(\frac{2\,\omega^{2}\,t}{\omega^{2}+1}-1\right)\cosh(\omega\tf)}{\omega\left(\tf -\frac{2}{\omega^{2}+1}\right)\sinh(\omega\,\tf)+2\, \left(\frac{\omega^{2}\,\tf}{\omega^{2}+1}-1\right)\cosh(\omega\,\tf)+2}\end{split}
 \end{align}
 and
 \begin{align}\begin{split}
 	\lambda_{t}=&x_{\mathscr{f}}\,\omega\,\frac{t\,\sinh(\omega\,\tf)-\sinh(\omega\,t)+\frac{\omega^{2}-1}{\omega^{2} +1 }\,\Big{(}\sinh(\omega\,\tf)-\sinh(\omega(\tf-t))\Big{)}}{\omega\left(\tf -\frac{2}{\omega^{2}+1}\right)\sinh(\omega\,\tf)+2\, \left(\frac{\omega^{2}\,\tf}{\omega^{2}+1}-1\right)\cosh(\omega\,\tf)+2} \\
 	&+x_{\mathscr{f}}\frac{1-\cosh(\omega\,t)+\frac{2\,\omega^{2}\,t}{\omega^{2}+1}\cosh(\omega\,\tf)+\frac{\omega^{2}-1}{\omega^{2} +1 }\,\Big{(}\cosh(\omega\,\tf)-\cosh(\omega(\tf-t))\Big{)}}{\omega\left(\tf -\frac{2}{\omega^{2}+1}\right)\sinh(\omega\,\tf)+2\, \left(\frac{\omega^{2}\,\tf}{\omega^{2}+1}-1\right)\cosh(\omega\,\tf)+2}
 	\end{split}
 \end{align}

\subsection{Conditional work minimisation}
 \label{app:hcmw}
 
 If, instead of (\ref{realworkbc}), we impose (\ref{naive:Lfinal}), we get
\begin{align}
	&	f_{t}^{\scriptscriptstyle{(1)}}=\frac{\omega\, t\, \sinh (\omega \tf)-\cosh (\omega \tf)-\cosh (\omega t)+\cosh (\omega (\tf-t))+1}{\omega \,\tf \,\sinh (\omega \tf)-2 \cosh(\omega \tf)+2}
	 \\[20pt]
	\begin{split}& f_{ t}^{\scriptscriptstyle{(2)}}=\frac{ \tf \cosh (\omega t)- t \cosh (\omega \tf)-(\tf-t)}{\omega \,\tf \,\sinh (\omega \tf)-2 \cosh(\omega \tf)+2} \\
    &\qquad\qquad\qquad\qquad+\frac{\Big{(}\sinh (\omega \tf)-\sinh (\omega (\tf-t))-\sinh (\omega t)\Big{)}}{\omega\bigl(\omega \,\tf \,\sinh (\omega \tf)-2 \cosh(\omega \tf)+2\bigr)}\end{split}
\end{align}
and
\begin{align}
	&	f_{t}^{\scriptscriptstyle{(3)}}=\omega\frac{\tf \,\sinh (\omega t)-t \,\sinh (\omega \tf)}{\omega\, \tf \,\sinh (\omega \tf)-2 \cosh(\omega \tf)+2} \\[20pt]
	& f_{ t}^{\scriptscriptstyle{(4)}}=\omega\frac{\sinh (\omega \tf)-\sinh (\omega t) -\sinh (\omega (\tf-t)) }{\omega\, \tf \,\sinh (\omega \tf)-2 \cosh(\omega \tf)+2}
\end{align}

 \subsection{Work minimisation over the system final state}
 \label{app:wmsv}
 
If we impose (\ref{naive:xinit}), (\ref{naive:Linitial}), (\ref{naive:Lfinal}) and (\ref{sl:wmsbc}) we obtain:
 \begin{align}\begin{split}
 	\mathscr{x}_{t}=&\ell\,\frac{\big{(}\omega \cosh \left(\omega t\right)-\sinh \left(\omega t\right)+\omega \left(t-1\right)\big{)}\cosh \left(\frac{\omega \tf }{2}\right)}{\left(\left(\tf +2\right)\omega ^{2}-2\right)\sinh \left(\frac{\omega \tf }{2}\right)+\tf \,\omega \cosh \left(\frac{\omega \tf }{2}\right)} \\
    &-\ell\,\frac{\big{(}1-t\omega ^{2}-\cosh \left(\omega t\right)+\omega \sinh \left(\omega t\right)\big{)}\sinh \left(\frac{\omega \tf }{2}\right)}{\left(\left(\tf +2\right)\omega ^{2}-2\right)\sinh \left(\frac{\omega \tf }{2}\right)+\tf \,\omega \cosh \left(\frac{\omega \tf }{2}\right)}
    \end{split}
 \end{align}
 and
 \begin{align}\begin{split}
 	\lambda_{t}=&\ell\,\frac{\big{(}\left(\omega ^{2}-1\right)\sinh \left(\omega t\right)+t\omega \big{)}\cosh \left(\frac{\omega \tf }{2}\right)}{\left(\left(\tf +2\right)\omega ^{2}-2\right)\sinh \left(\frac{\omega \tf }{2}\right)+\tf \,\omega \cosh \left(\frac{\omega \tf }{2}\right)} \\
    &-\ell\,\frac{\big{(}\left(\omega ^{2}-1\right)\cosh \left(\omega t\right)-\left(t+1\right)\omega ^{2}+1\big{)}\sinh \left(\frac{\omega \tf }{2}\right)}{\left(\left(\tf +2\right)\omega ^{2}-2\right)\sinh \left(\frac{\omega \tf }{2}\right)+\tf \,\omega \cosh \left(\frac{\omega \tf }{2}\right)}
        \end{split}
 \end{align}

 \section*{References}	
\bibliography{work} 
\bibliographystyle{iopart-num}
\end{document}